\documentclass[onecolumn,aps,preprint,superscriptaddress]{revtex4}
\usepackage{epsfig}
\newcommand{\be}{\begin{equation}}
\newcommand{\ee}{\end{equation}}
\newcommand{\bea}{\begin{eqnarray}}
\newcommand{\eea}{\end{eqnarray}}

\newcommand{\bc}{\begin{center}}
\newcommand{\ec}{\end{center}}

\newcommand{\forget}[1]{}

\begin{document}

\title{Performance of Magnetic Quantum Cellular Automata and Limitations due to Thermal Noise}

\author{Federico M. Spedalieri}
\email{federico@ee.ucla.edu}
\affiliation{Department of Electrical Engineering, University
of California, Los Angeles, Los Angeles, California 90095, USA}
\author{Ajey P. Jacob}
\affiliation{Technology Strategy, Technology and Manufacturing Group, Intel Corporation,
2200 Mission College Blvd., Santa Clara, California 95052, USA}
\author{Dmitri Nikonov}
\affiliation{Technology Strategy, Technology and Manufacturing Group, Intel Corporation,
2200 Mission College Blvd., Santa Clara, California 95052, USA}
\author{Vwani P. Roychowdhury}
\affiliation{Department of Electrical Engineering, University
of California, Los Angeles, Los Angeles, California 90095, USA}

\date{\today}

\begin{abstract}
Operation parameters of magnetic quantum cellular automata are evaluated
for the purposes of reliable logic operation. The dynamics of the nanomagnets is simulated 
via the Landau-Lifshitz-Gilbert equations with a stochastic magnetic field corresponding to thermal 
fluctuations. It is found that in the macrospin approximation the switching speed 
does not change under scaling of both size and distances between nanomagnets.
Thermal fluctuations put a limitation on the size of nanomagnets, since 
the gate error rate becomes excessive for nanomagnets smaller than 200nm at room temperature.
\end{abstract}

\maketitle

\section{Introduction}

The success of computing in the past 40 years was based on scaling the complementary metal-oxide-semiconductor (CMOS) transistors to the nanoscale size 
\cite{ITRS2007}. As it is anticipated that this scaling will approach limits
defined by the quantum theory and thermodynamics \cite{Zhirnov2003}, the 
search is on for alternative logic technologies \cite{bourianoff2007,hutchby2008},
which would be able to supplement CMOS and have certain advantages compared to it.
One promising technology among them is spintronics and nanomagnetics
\cite{zutic2004}.

Magnetic Quantum Cellular Automata (MQCA) have been proposed as one of the types of spintronic
logic.
MQCA are based on bistable nanomagnet elements that can perform basic logic
operations by means of magnetostatic interactions.
Nanomagnets are typically arranged in the shape of crosses - majority gates.
A majority gate has three inputs and one output. The output's logic state is determined
by the 'majority voting' of the logic states of the inputs.
This gate is naturally suited
for the magnetic dipole-dipole interaction that is the basis
of MQCA. It also allows us to perform AND and OR logical functions by
fixing one of the inputs, and (in combination with the NOT element) 
it can be used to perform
any logical operation.
Another type of spintronics - domain wall logic \cite{allwood2005}
can also be rendered in the form of majority gates \cite{nikonov2008}.
A chain of nanomagnets carrying the logic variables was demonstrated
by Cowburn and Welland \cite{cowburn2000a}.
Later, a majority gate based
on these principles was proposed and experimentally implemented~\cite{imre2006a}. 

To be a viable alternative to CMOS logic, MQCA must show that 
they can achieve a better (or at least similar) performance level
at least in one of the benchmarks, 
such as size, speed, switching energy, bit stability
and scalability.
Some of these issues have been studied through 
simulations~\cite{parish2003a,csaba2002a}.
In this paper, our goal is to estimate how far can we push the limits of
MQCA performance for all the benchmarks presented above. To this end, we will
analyze a simplified model of MQCA that captures the basic physical principles that
govern its behavior. We pay a special attention to the limitation stemming from the
thermal fluctuations of the magnetization.

The paper is organized as follows. In Section \ref{size} we
show how the bit stability of an MQCA element puts a lower bound on its size.
In Section \ref{dynamics} we introduce a simple model of the MQCA dynamics and
use it to simulate the behavior of an MQCA majority
gate and study the speed of a signal propagating
along a chain of nanomagnets. In Section \ref{initial} we discuss the relationship
between MQCA initialization and its stability. In Section \ref{thermal} we simulate the 
effects of thermal fluctuations and study their impact on the error rate of the majority
gate. Finally, in Section \ref{conclusions}
we summarize our results and present our conclusions.  

\section{bit stability and minimum size}
\label{size}

Our first step will be to study what type of constraints bit stability imposes on the 
size of MQCA. The basic element of MQCA is a nanomagnet that is used to store
a single bit of information. Usually the nanomagnets are elongated along some direction 
which determines the easy axis of magnetization due to shape anisotropy.
This bit is represented by the magnetization
direction of this nanomagnet: ``0'' for the magnetization 'pointing up', 
i.e., along positive easy axis, and ``1''
for the magnetization 'pointing down', i.e., along negative easy axis. 
We thus need to require these two configurations
to be stable and separated by an energy barrier to prevent bit-flip errors. 
Even though material properties such as the uniaxial anisotropy can be exploited to
produce such a bistable system, shape anisotropies are more advantageous
to produce such a result, and most proposals of MQCA are essentially based
on this idea. 

Our mathematical model is based on the free energy of a nanomagnet with
uniform magnetization $\mathbf{M}$. It includes contributions from the shape anisotropy,
material anisotropy, and the energy in the external magnetic field (see~\cite{d'aquino2004a} 
for a derivation).
\be
\label{staticenergy}
E = K_1 (1- (\mathbf{m}.\mathbf{\hat{e}}_{axis})^2) V +\frac{1}{2} \mu_0 M_s^2 V 
\mathbf{m}.\mathbf{\cal{N}}.\mathbf{m} - \mu_0 M_s V \mathbf{m}.\mathbf{H}_{ext},
\ee
where $\mathbf{m}=\frac{\mathbf{M}}{M_s}$ is the normalized magnetization
(note that $|\mathbf{m}|=1$); $M_s$ is the saturation magnetization of the material;
$V$ is the volume of the nanomagnet; $\mu_0$ is the permeability of vacuum; $K_1$ is the
uniaxial anisotropy of the material and $\mathbf{\hat{e}}_{axis}$ is a unit vector 
in the direction of the easy axis; $\mathbf{\cal{N}}$ is the demagnetizing tensor;
and $\mathbf{H}_{ext}$ is the external field. 
The demagnetizing tensor can be diagonalized by finding its principal axes, 
and its diagonal elements
are positive and satisfy $N_x + N_y + N_z =1$.
We will consider that our nanomagnet is
a rectangular prism whose symmetry axes are aligned with the cartesian axes. 
We will also assume that the 
easy axis of the crystalline uniaxial anisotropy is aligned with the $y$ axis.
The explicit expression for 
these demagnetizing factors can be found in~\cite{aharoni1998a}.

Let us consider the case of a vanishing external field.
If $a,b$ and $c$ are the dimensions of the nanomagnet in the $x, y$ and $z$ directions,
we will assume that $b>a>c$, which corresponds to a rectangular prism elongated
in the $y$ direction. This choice of proportions translates into an
inverse ordering of the demagnetizing factors ($N_z > N_x > N_y$). This makes the
$z$ direction the least energetically favorable. It is easy to see that 
the energy is minimal when the magnetization points in the $y$ direction, either
up or down. These are the two stable states that encode a bit of information.
Then the energy barrier between
these two minima is smaller when
we consider the magnetization to be in the $x-y$ plane. To compute this
energy barrier, we just need to evaluate (\ref{staticenergy}) in the $x$
and $y$ directions and subtract them. Then we have
\be
\label{energybarrier}
\Delta E = E(\mathbf{m}=\mathbf{\hat{e}}_x)-E(\mathbf{m}=\mathbf{\hat{e}}_y)=
\frac{1}{2} \mu_0 M_s^2 V \left[ N_x -(N_y - \frac{2 K_1}{\mu_0 M_s^2})\right].
\ee
From this equation we can extract a few useful facts: ({\it i}) the energy scale
is given by $\frac{1}{2} \mu_0 M_s^2 V $; ({\it ii}) the energy barrier, and hence the energy
dissipation, scales down with the volume of the nanomagnet; ({\it iii}) the geometrical
anisotropy can be used to control the height of the barrier; ({\it iv}) the 
uniaxial crystal anisotropy can be seen as a correction to the geometrical anisotropy.

The height of the energy barrier will determine the stability of the information stored
in the nanomagnets, and hence its bit stability. 
The thermal fluctuations will cause the direction of the magnetization to 
vary and with a certain probability to turn over 90 degrees - the direction of the energy
saddle point. After that the magnetization will flip to the other energy minimum.
In a simple model the probability of the nanomagnet's magnetization flipping its direction 
due to thermal noise is given by $p_{flip} = \exp (-\Delta E / k_B T)$,
where $k_B$ is the Boltzmann constant.
Since we are interested in MQCA as an alternative to CMOS-based logic, it is natural to require
this error probability to be at least of the same order as that of CMOS transistors,
which is of the order of $10^{-17}$.
This corresponds to the condition $\Delta E/k_BT > 40$.
For room temperature we have $k_B T \approx 0.026 eV$, and so we need
$\Delta E \approx 1 eV$ or larger. 
The energy barrier height gives an approximate estimate of the energy that will be
dissipated every time we switch the magnetization direction of a nanomagnet.
The exception would be slow adiabatic switching regime \cite{behinaein2008} 
which we do not consider here.

The lower bound on the height of the energy barrier, coupled with equation (\ref{energybarrier})
allows us to extract a lower bound on the size of the nanomagnets. Since the energy barrier
depends on the volume of the nanomagnet, any lower bound on it will translate
into a lower bound on the volume.  Assuming that the geometrical 
anisotropy is due to a 2:1 aspect ratio between the length and
width of the prism, we can plot the values of thickness and width that are
required to obtain a $1 eV$ energy barrier. In Figure 1, we present this plot
for three different materials: permalloy, CoFeB and Fe (with saturation magnetizations
equal to $800 kA/m$, $1180 kA/m$ and $1750 kA/m$, respectively.)
\begin{figure}[ht]
\centerline{\includegraphics[scale=0.7]{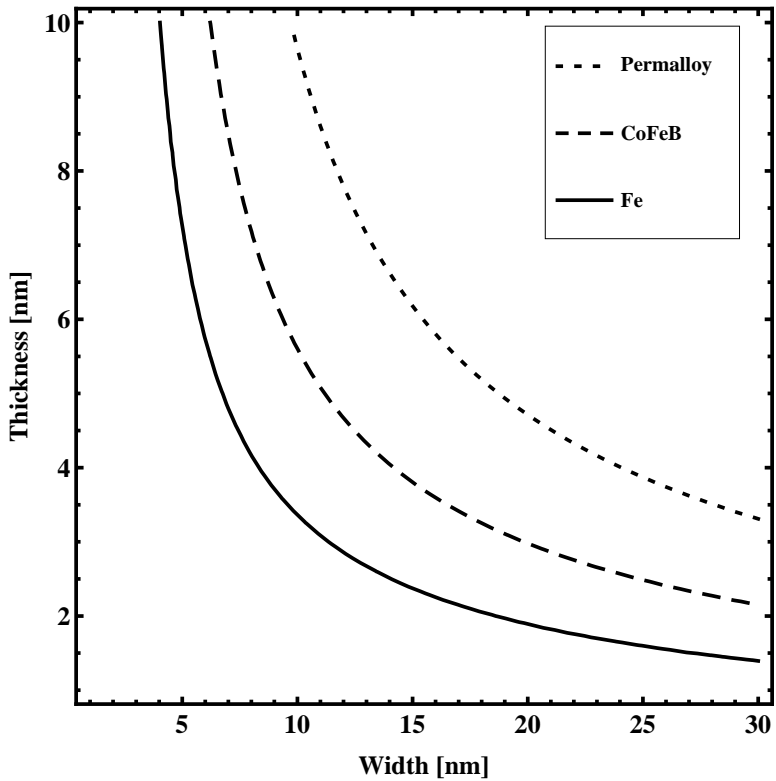}}
\caption{\label{Fig:minsize} Thickness vs. width for a nanomagnet with an energy barrier
of $1 eV$. The length is taken to be twice the width.}
\end{figure}
For example, in permalloy, we can see that for a thickness of $6nm$, the nanomagnet needs to have 
a $15 nm$ width and a $30 nm$ length. Clearly, there is an advantage for higher values
of the saturation magnetization, since we can achieve the same energy barrier height
with a smaller volume (see Eq. (\ref{energybarrier}).)

\begin{figure}[ht]
\centerline{\includegraphics[scale=0.7]{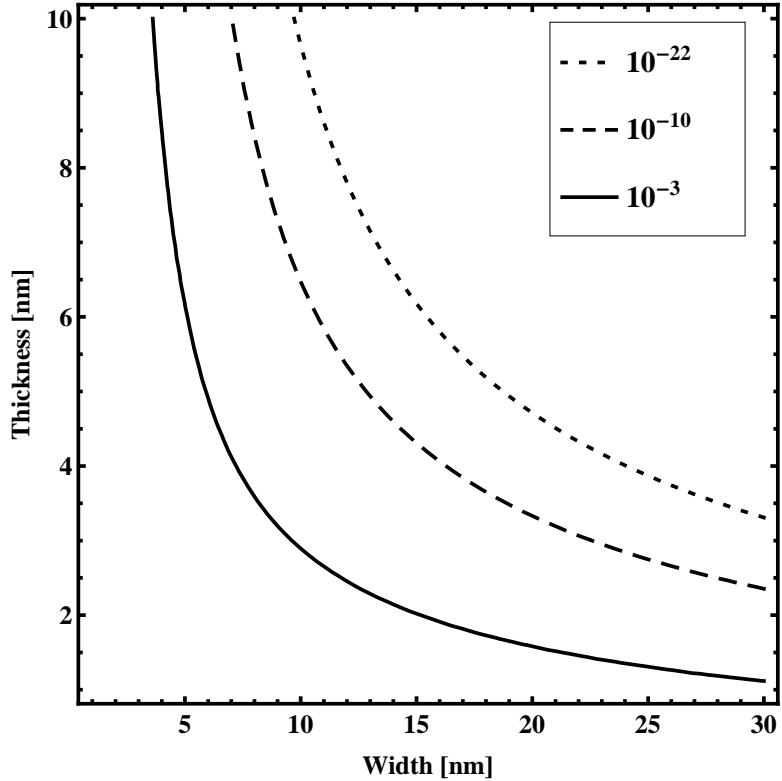}}
\caption{\label{Fig:Perror} Thickness vs. width for different values of the error probability 
(plot corresponds to permalloy, and a 2:1 aspect ratio.)}
\end{figure}

It can be argued that the very high bit stability we are requiring (error rate $\simeq 10^{-17}$)
might be appropriate for a memory device, but may not need to be that high for a logic
device. For MQCA, we only need the nanomagnets to maintain their state only during
the time it takes to perform a certain computation. We might be able to 
reduce the size even further if we somewhat relax the bit stability requirements. However,
given that the dependence of the error probability with the energy barrier is 
exponential, a small reduction in size can have a huge impact on the bit stability.
We can illustrate this point by repeating the plot in Figure 1 for permalloy,
but for different values of the error probability (Figure \ref{Fig:Perror}.)
We can see the rapid increase of the error probability even for a modest reduction on
the size of the nanomagnet. This shows that the
lower limit on the size of MQCA is a rather strong one if we want to preserve
bit stability.  

\section{Dynamics and speed}
\label{dynamics}

To estimate the speed of MQCA-based logic devices we will simulate their
behavior using the Landau-Lifshitz-Gilbert (LLG) equations~\cite{landau1935a,gilbert1955a}. 
Since we are only 
interested in an order of magnitude estimate, we will skip the detailed micromagnetic
simulations that are usually discussed in the literature 
\cite{e2008}, and instead work with a very
simple model of the MQCA. We will model each nanomagnet as a macrospin, but we will 
include the effects of geometrical and crystalline anisotropies
in the computation of the effective field. This approximation is equivalent to assuming the
magnetization is uniform over the whole volume of a nanomagnet at any time, and 
neglecting magnetic moments higher than the dipole moment \cite{csaba2002a,csaba2002b}. 
We expect this approximation
to improve for decreasing nanomagnet size, since the exchange interaction tends to force
the magnetization to be uniform on a length scale of about $10 nm$. From our discussion
in the previous section, we are interested in nanomagnet sizes of the order of tens of nanometers,
so we are not that far from that regime. In any case, we are interested in an upper
bound for the speed of MQCA-based logic, and a full simulation will most 
likely produce a slower device.

The LLG equations \cite{fidler2000,miltat2002} for the macrospin model are
\be
\label{LLG}  
\frac{d \mathbf{M}^{(i)}}{d t} = -\frac{\gamma}{1+\alpha^2} \mathbf{M}^{(i)} \times 
\mathbf{H}^{(i)}_{eff} - \frac{\gamma \alpha}{(1+\alpha^2) M_s} 
 \mathbf{M}^{(i)} \times (\mathbf{M}^{(i)} \times \mathbf{H}^{(i)}_{eff} ),
\ee
where $\mathbf{M}^{(i)}$ is the magnetization of the $i^{th}$ nanomagnet, 
$\mathbf{H}^{(i)}_{eff}$ is the effective field at the position of the
$i^{th}$ nanomagnet, $\gamma = g |e| / 2 m_e c = 2.21 \times 10^5 m A^{-1} sec^{-1}$ is the 
Lande factor, and $\alpha$ is the Gilbert damping constant, which depends on the material
and the environment of the nanomagnet
and typically has values in the range $0.001\,- \, 0.1$. The effective field includes the
contributions of any external field, the nanomagnet self-field and the field due
to the dipole-dipole interaction with other nanomagnets.
\be
\label{Heff}
\mathbf{H}^{(i)}_{eff} = \mathbf{H}^{(i)}_{ext} - \mathbf{\cal{N}} \cdot \mathbf{M}^{(i)} +
\sum_j \mathbf{C}_{(ij)} \mathbf{M}^{(j)}.
\ee
In this expression we are assuming that all nanomagnets have the same shape,
and hence the demagnetizing tensor $\mathbf{\cal{N}}$ is the same for
all nanomagnets. This term can also include the effects of uniaxial
crystalline anisotropy if we redefine the corresponding
demagnetizing factor $N_y \rightarrow N_y - \frac{2 K_1}{\mu_0 M_s^2}$, where 
$y$ is the easy axis of the crystalline anisotropy.
The last term on the RHS of (\ref{Heff}) represents
the dipole-dipole interaction between nanomagnets, and the matrices $\mathbf{C}_{(ij)}$
are coupling constants determined by their size and relative positions.
If $(x^{(i)},y^{(i)},z^{(i)})$ are the coordinates
of the $i^{th}$ nanomagnet, we define 
the coordinate differences for a pair of nanomagnets as
$d_x^{(ij)} =  x^{(i)} - x^{(j)}$, $d_y^{(ij)} =  y^{(i)} - y^{(j)}$,
$d_z^{(ij)} =  z^{(i)} - z^{(j)}$, and 
the distance between nanomagnet centers as
$d^{(ij)} = \sqrt{(d_x^{(ij)})^2 + (d_y^{(ij)})^2 +(d_x^{(ij)})^2}$.
The coupling constant matrices $\mathbf{C}_{(ij)}$ are given by
\be
\label{C}
\mathbf{C}_{(ij)} = \frac{V^{(j)}}{4 \pi (d^{(ij)})^5} \left(\begin{array}{ccc}
3 (d_x^{(ij)})^2 - (d^{(ij)})^2 & 3 d_y^{(ij)} d_x^{(ij)} & 3 d_z^{(ij)} d_x^{(ij)}  \\
 3 d_x^{(ij)} d_y^{(ij)} & 3 (d_y^{(ij)})^2 - (d^{(ij)})^2 & 3 d_z^{(ij)} d_y^{(ij)}  \\
 3 d_x^{(ij)} d_z^{(ij)} &  3 d_y^{(ij)} d_z^{(ij)} & 3 (d_z^{(ij)})^2 - (d^{(ij)})^2   \\
\end{array} \right).
\ee
In our case, since all nanomagnets will be in the $(x,y)$ plane, this
expression simplifies since $d_z^{(ij)} = 0$. An important fact about
the matrices $\mathbf{C}_{(ij)}$ is that they are dimensionless, and hence
invariant under scaling of both the sizes of nanomagnets and 
the distances between nanomagnets. 
We will see that this property is
preserved by the LLG equations in our model.  

To simplify the simulation and analysis it is useful to normalize the
LLG equations. This is accomplished using the following definitions
\bea
\label{redefinitions}
\mathbf{m}^{(i)} &=& \frac{\mathbf{M}^{(i)}}{M_s} \nonumber \\
\mathbf{h}_{eff}^{(i)} &=& \frac{\mathbf{H}_{eff}^{(i)}}{M_s} \nonumber \\
t' &=& t (\gamma M_s),
\eea
where now all the quantities on the LHS of (\ref{redefinitions}) are
dimensionless (note that $[\gamma M_s] = sec^{-1}$.) With these rescalings
and using vector identities and the obvious fact that 
$\frac{d \mathbf{M}^{(i)}}{d t} . \mathbf{M}^{(i)}=0$, we can rewrite
the normalized LLG equations in an implicit form that simplifies
the implementation of the simulation,
\be
\label{normLLG}
\frac{d \mathbf{m}^{(i)}}{d t'} = -\mathbf{m}^{(i)} \times 
\mathbf{h}^{(i)}_{eff} + \alpha 
 \mathbf{m}^{(i)} \times \frac{d \mathbf{m}^{(i)}}{d t'}.
\ee
These equations have the property that the value of the magnetization is constant,
$|\mathbf{m}^{(i)}(t')| \equiv 1, \,\forall t'$,
and this feature must be preserved in the discretized numerical model.
To do this we employ
the mid-point method~\cite{d'aquino2006a} with which this constraint is automatically satisfied.

\begin{figure}[ht]
\centerline{\includegraphics[scale=0.5]{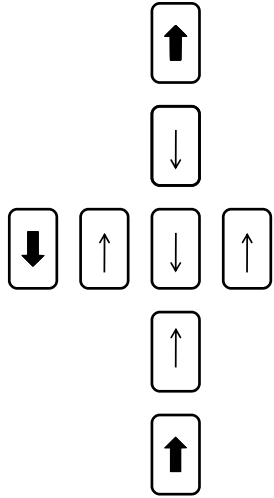}}
\caption{\label{Fig:maj} Majority gate: the thick arrows represent nanomagnets with fixed
magnetization that simulate inputs to the gate. The remaining nanomagnet align their
magnetization in order to minimize the energy of the system from an initial magnetization
in the $x$ (horizontal) direction. The output of the gates can be extracted from the
magnetization of the ``output'' nanomagnet on the right.}
\end{figure}

To estimate the speed with which MQCA switch, we simulated the
behavior of the majority gate. 
Let us first briefly review its operation. The nanomagnets
forming the gate are arranged as seen in Figure \ref{Fig:maj}. 
We also
include three nanomagnets with fixed magnetization that are used
to simulate the inputs of the gate. The nanomagnets that form the
gate are initially magnetized in the $x$ direction, and then are left
to evolve driven by the magnetic dipole interaction. The magnetization
of each nanomagnet will tend to align itself with the field
produced by the other nanomagnets at its position. The geometric anisotropy
will force the magnetization to lay in the $y$ direction, and the
influence of other nanomagnets will decide if it ends
pointing up or down. The fields of the three inputs will add 
at the position of the central magnet and decide its direction
of magnetization, hence computing the majority of the input signals.
Finally, this signal can be read on the output magnet. Note that a signal
that propagates horizontally is inverted every time it is received
by the next nanomagnet (due to the antiferromagnetic coupling). This
does not affect the function of the gate, though this feature must be
tracked in order to correctly interpret the output of any MQCA-based gate.

Again, in order to extract numerical estimates from the simulation, we specified
the properties of the material ($M_s$ and $K_1$) to be those
of permalloy. The value of the Gilbert damping constant did not have 
a big effect on the simulation when confined to the typical range 
$0.001 - 0.01$. We found that the typical gate time, measured as
the time it took the output to reach 90\% of its final magnetization,
was about $700 ps$. An interesting feature of our model is
that the normalized equations (\ref{normLLG}) are invariant under
changes of scale, which means that \emph{the gate time is
independent of size}. Even though this is only true in this simplified
model, and making less approximations will likely break this invariance,
whatever effects this may have on the the gate time will likely be 
of higher order. 
This is in contrast to CMOS logic \cite{ITRS2007} as well as 
MQCA based on magnetic wires (rather than discrete nanomagnets)
\cite{nikonov2008}.

From the form of the normalized equations we can see that the
speed of this gate will depend on the material properties.
In particular, the speed of the gate increases linearly with 
the saturation magnetization of the material. This follows from the 
scaling of actual time $t$ with respect to normalized time $t'$, as defined
by (\ref{redefinitions}).

\begin{figure}[ht]
\centerline{\includegraphics[scale=0.75]{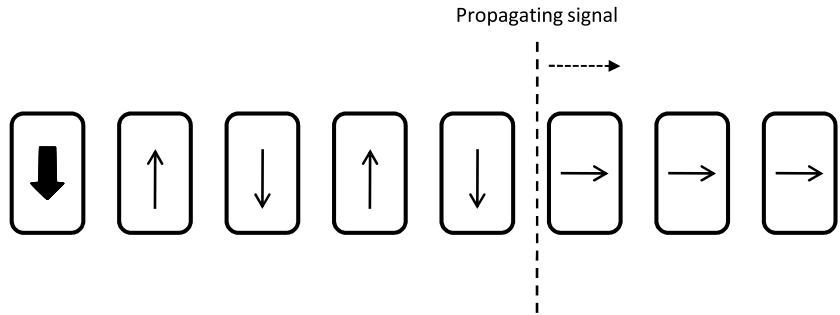}}
\caption{\label{Fig:horwire} Signal propagation through a horizontal wire made up of a chain
of nanomagnets. The antiferromagnetic coupling forces neighboring nanomagnets
to become antiparallel.}
\end{figure}

Another issue that needs to be considered when analyzing
the speed of MQCA-based information processing, is the 
speed of propagation of information. In MQCA this is
accomplished by chains of nanomagnets that are initially
magnetized in the $x$ direction, and evolve according
to the dipole-dipole interaction propagating a signal,
as can be seen in Figure \ref{Fig:horwire} for the case
of a horizontal wire. 
Note that the antiferromagnetic
coupling forces neighboring nanomagnets to be antiparallel.
For vertical wires the coupling is ferromagnetic and the
nanomagnets magnetization tends to become parallel. 

This evolution follows the same dynamical equations presented
in the previous section, so we can use them to simulate
the propagation of a signal along a chain of nanomagnets and
estimate its speed. For nanomagnets made of permalloy with a width of 
about $10 nm$, separated by $15 nm$, the speed of signal 
propagation is around $100 m/sec$, or equivalently, $150 psec$ per magnet.
This is of the order of
the speed of sound, and would certainly limit the speed
of an integrated MQCA chip if communication is done using
the same principles as logic. This speed depends on the material
through the saturation magnetization, but only linearly,
so it is not likely that choosing a different material
will solve this problem for MQCA.

\section{Initialization and bit stability}
\label{initial}

\begin{figure}[ht]
\centerline{\includegraphics[scale=0.9]{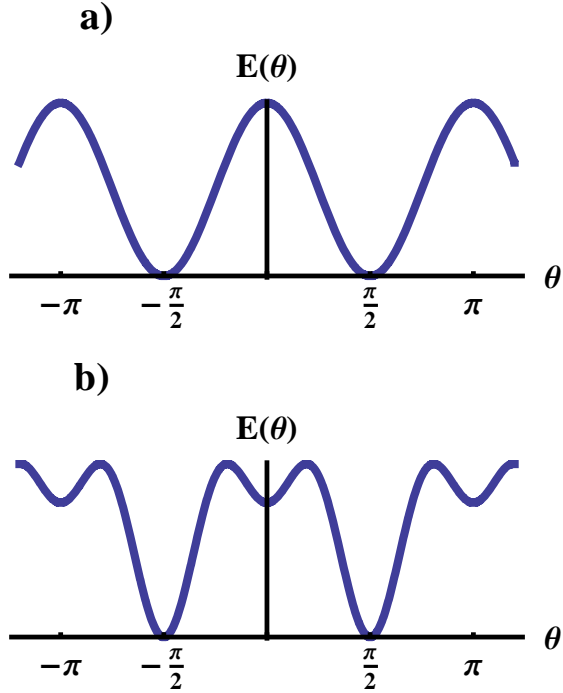}}
\caption{\label{Fig:energyprofile} a) Energy profile for geometrical and uniaxial anisotropies. Stable
configurations correspond to magnetization in the $y$ direction (up or down). Magnetization in the
$x$ direction (initial configuration) is an unstable equilibrium point. b) Including a biaxial
anisotropy produces local minima for magnetization in the $x$ direction, stabilizing the 
initial configuration.}
\end{figure}

As discussed before, in order to run a MQCA-based logic gate it is
necessary to initialize the magnetization of all nanomagnets
in the $x$ direction (i.e., the hard axis.) In terms of energy, this
corresponds to placing all nanomagnets at the top of the energy
barrier created by the geometrical and crystalline anisotropies (see 
Figure \ref{Fig:energyprofile} a).)
However, this configuration corresponds to an unstable equilibrium
point for each nanomagnet, and it should be expected that small
perturbations due to thermal effects and stray fields will randomly force the
nanomagnets to relax to one of their stable configurations
independent from the input signals. 

This is an important issue for any implementation of MQCA-based logic and
some possible solutions have been suggested. One consists of exploiting
the biaxial anisotropy of the material to create a stable configuration
around the initialization direction, by generating a local minimum of
the energy~\cite{carlton2008}. If we consider the magnetization confined to the 
$x-y$ plane, and note as $\theta$ the angle between the magnetization
direction and the $x$ axis, the geometric and uniaxial anisotropy
result in an energy profile proportional to $\cos^2 (\theta)$ as
can be seen in Figure \ref{Fig:energyprofile} a). The biaxial anisotropy
introduces another term that is proportional to $\sin^2 (2\theta)$, and
by carefully choosing the parameters we can produce a local minimum
for $\theta = 0$, as seen in Figure \ref{Fig:energyprofile} b). 

This energy minimum provides a latch mechanism that keeps the initialized 
nanomagnets pointing in the $x$ direction while the information from the 
input signal propagates through the chain of magnets. Once again, the 
effectiveness of this local minimum to trap the magnetization
direction against thermal fluctuations, will depend on the height
of the energy barrier around it (i.e., the energy difference
between the peaks and the local minimum in Figure \ref{Fig:energyprofile} b).)
The reasoning of Section \ref{size} applies to estimate the energy
of this barrier necessary to preserve the bit in its local energy minimum
for sufficiently long time, and hence obtain an estimate of the strength of the
required biaxial anisotropy. 
We realize that the requirements to the height of this barrier are contradictory -
it should be high enough to prevent spontaneous transition to one of the 
global minima before the signal reaches the bit; it also needs to be low enough so that
the signal can reliably switch it to the desired local minimum.
In the next section we 
simulate the behavior of the majority gate, including the biaxial anisotropy,
in the presence of thermal fluctuations. 

\section{Thermal effects and gate error probability}
\label{thermal}

In this section we model the effects of the thermal fluctuations on the operation of 
MQCA. We especially focus on gate errors caused by spontaneous transitions from the local 
energy minimum after the initialization of elements of MQCA.

Our simulations will use the stochastic LLG equations based on the midpoint rule derived
by d'Aquino {\it et al.} in~\cite{d'aquino2006a}. The only difference with the above model
(Section \ref{dynamics})
will be the inclusion of an extra term that represents the field generated
by the biaxial anisotropy (we show in the appendix that the introduction
of this term does not affect the useful properties of the discretized
equations.)

Let us start by considering the extra term in the normalized effective field 
that is responsible for the biaxial anisotropy acting on nanomagnet $(i)$,
\be
\label{biaxialfield}
\mathbf{h}^{(i)}_{eff(biaxial)} = -\frac{2 K_2}{\mu_0 M_s^2} 
\left( m_x^{(i)} (1-(m_x^{(i)})^2) \mathbf{\hat{x}} +  m_y^{(i)} (1-(m_y^{(i)})^2) \mathbf{\hat{y}} +
m_z^{(i)} (1-(m_z^{(i)})^2) \mathbf{\hat{z}} \right).
\ee
The biaxial anisotropy constant $K_2$ has dimensions of $J m^{-3}$. It is not difficult to show that,
when restricted to the $x-y$ plane, the contribution to the energy of this term is proportional
to $\sin^2 (2\theta)$, with $\theta$ the angle between the magnetization direction and
the $x$ axis. In order to have a local minimum around $\theta = 0$, the constant $K_2$
must satisfy the condition
\be
\label{K2min}
K_2 > K_{2min} = \frac{1}{2} \mu_0 M_s^2 V \left[ N_x -(N_y - \frac{2 K_1}{\mu_0 M_s^2})\right].
\ee

The thermal fluctuations manifest themselves as random variations of the overall 
magnetization of the nanomagnet.
We describe this process by the stochastic LLG equations
\cite{bertram2002,safonov2005}, which
 are obtained by adding a random force, or, in other words, a stochastic thermal 
 magnetic field $\mathbf{h}^{(i)}_{T}(t)$ 
to the effective field in (\ref{normLLG}). 
Note that we are considering a different 
thermal field for each nanomagnet, 
since it is usually assumed that the thermal fluctuations in different nanomagnets are uncorrelated. The random 
thermal field $\mathbf{h}^{(i)}_{T}(t)$ is assumed to be an isotropic vector Gaussian 
white-noise process with variance $\nu^2$, and so it can be expressed in terms of the 
Wiener process as $\mathbf{h}^{(i)}_{T}(t) dt = \nu \, d\mathbf{W}^{(i)}.$
Then, the stochastic LLG equations take the form
\be
\label{thermalLLG}
d \mathbf{m}^{(i)} = -\mathbf{m}^{(i)} \times \left(
\mathbf{h}^{(i)}_{eff} + \mathbf{h}^{(i)}_{eff(biaxial)}\right) dt -\mathbf{m}^{(i)} \times \nu \, 
d\mathbf{W}^{(i)} + \alpha\, 
 \mathbf{m}^{(i)} \times d \mathbf{m}^{(i)}.
\ee
The value of $\nu$ can be obtained from the fluctuation-dissipation theorem in thermal
equilibrium, and is given by $\nu = \sqrt{\frac{2 \alpha k_B T}{\mu_0 M_s^2 V}}$.
 
\begin{figure}[ht]
\centerline{\includegraphics[scale=0.7]{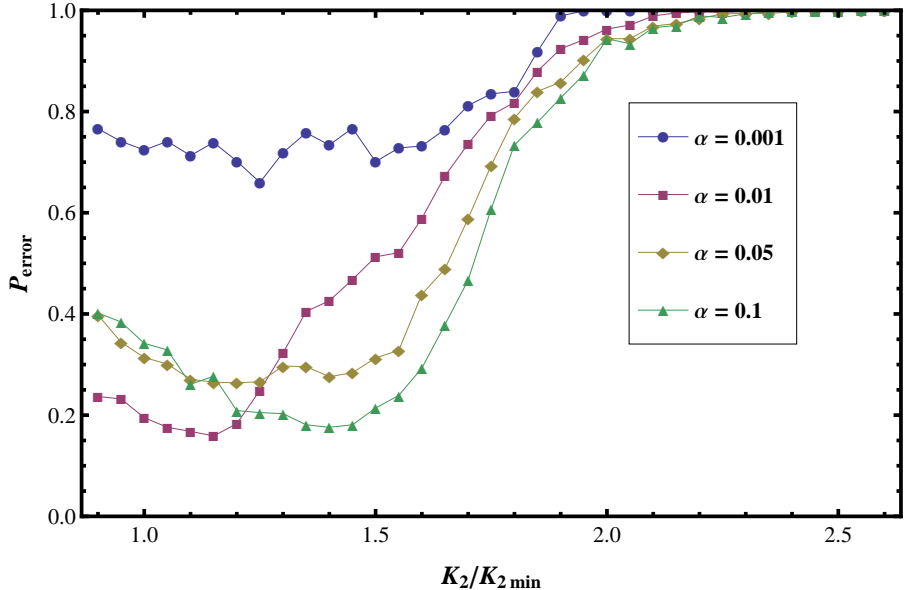}}
\caption{\label{Fig:sims}Error probability of the majority gate as a function
of the (scaled) biaxial anisotropy for different values of the damping constant ($T=300K$).}
\end{figure}

Using (\ref{thermalLLG}), we simulated the behavior of the majority gate for various 
values of size, 
damping constant, and temperature.
We fixed the
saturation magnetization and uniaxial anisotropy to be those of permalloy, and studied
the error rate of the gate as a function of $K_2$ and for several values of the 
damping constant $\alpha$. Starting with the nanomagnets initialized with
magnetization in the $x$ direction, each run simulated the evolution of the gate for
$2000 \, ps$. We considered the gate to be successful if the average of the output magnet
during the last $300 \, ps$ was larger than $80\%$ of the ideal output value (all runs
used the same set of fixed inputs.) In any other case, we considered that the gate failed.
For each value of the parameters $K_2$ and $\alpha$, we ran 1000 instances of the simulation.
The results are presented in Figure \ref{Fig:sims}.  
The error probability is plotted against the ratio of $K_2$ to $K_{2min}$, where
$K_{2min}$ is the minimum value of the biaxial anisotropy that produces a local
energy minimum around $\theta = 0$. If we increase $K_2$, we expect the error probability
to decrease when we pass $K_2 /K_{2min} =1$, since the biaxial anisotropy becomes more effective
in preventing a premature flipping of the nanomagnets spurred by the thermal fluctuations.
On the other hand, if we increase the biaxial anisotropy too much, the local energy minimum
is too deep for the signal to force the nanomagnet to flip. This is the behavior we
can appreciate in Figure \ref{Fig:sims}. For $K_2 > 2 K_{2min}$, the gate becomes essentially frozen
by the biaxial anisotropy; for $K_{2min} < K_2 < 2 K_{2min}$, the error probability seems 
to have a minimum for a certain value of $K_2$, that depends on the damping constant.
However, an important result of these simulations is that, for the particular 
temperature and size considered ($30nm \times 15nm \times 6nm$ magnets), 
the gate error rate exceeds a certain minimum value, $15\%$ in this case. 
The stabilizing effects of the biaxial anisotropy are either too weak, and spontaneous gate errors happen, or too strong, so that it prevents the normal evolution of the gate.

\begin{figure}[ht]
\centerline{\includegraphics[scale=0.7]{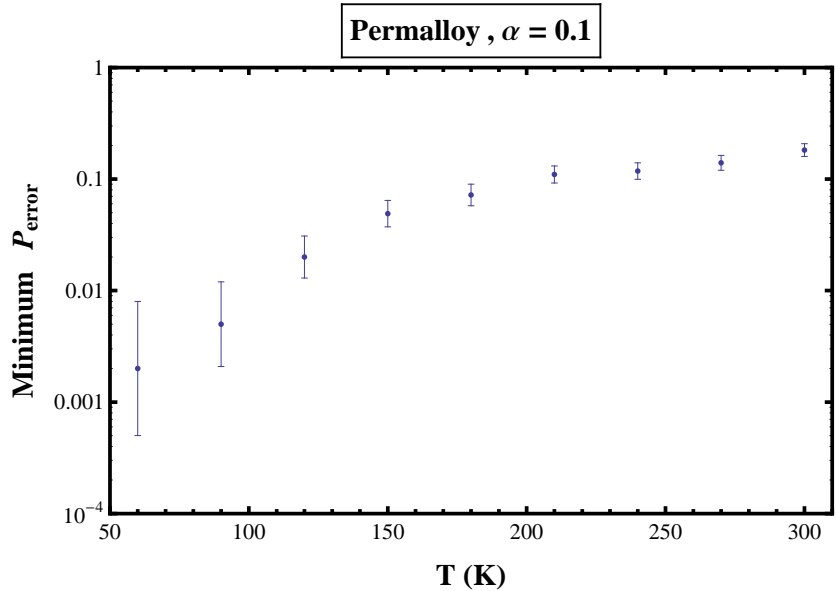}}
\caption{\label{Fig:temp}Minimum error probability of the majority gate as a function
of temperature.}
\end{figure}

One possible solution for the gate error probability will be to decrease
the temperature. Then, thermal fluctuations will be weaker and smaller values of the 
biaxial anisotropy will be enough to keep the magnets magnetized in the $x$ direction
until the signal, in the form of the magnetization of a neighboring magnet in the $y$ 
direction, reaches the magnet and makes it flip up or down. And since the required
biaxial anisotropy is not too large, it does not freeze the magnet in its initial
magnetization direction. We used our model to study the dependence of the gate error
probability
on the temperature, again running 1000 simulations for each value of the temperature and
the biaxial anisotropy, and then finding the minimum value of the error probability
for each temperature. These results are presented in Figure \ref{Fig:temp}.

We can see that, as expected, the error probability decreases with decreasing temperature, 
although this decrease seems rather slow for temperatures above $150K$. 
For temperatures below $30K$,
the error probability is below $0.001$, but it could not be accurately estimated
with the same number of simulation runs.

\begin{figure}[ht]
\centerline{\includegraphics[scale=0.7]{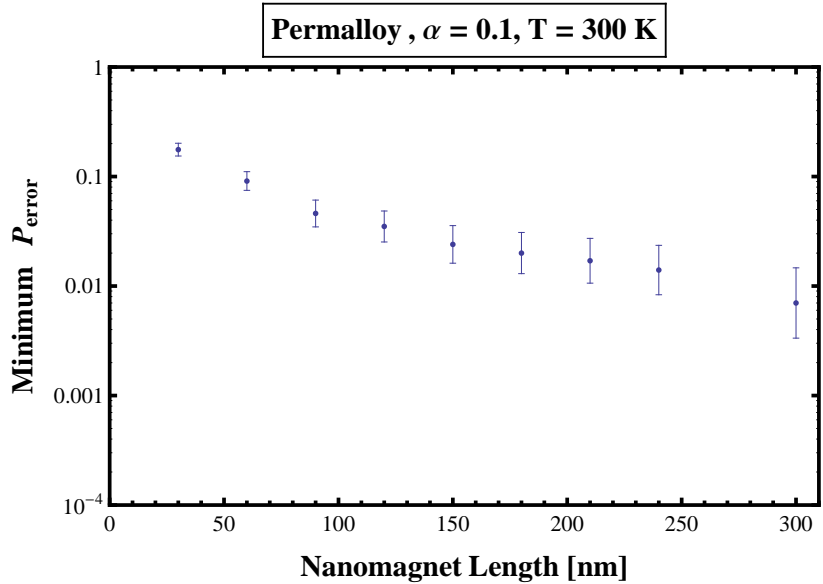}}
\caption{\label{Fig:MinErrorSize} Minimum error probability of the majority gate as a function
of the length of the nanomagnets.}
\end{figure}

Another approach to lowering the error probability of the gate is to increase the
size of the magnets. We know that larger magnets have a larger energy barrier between
the states of up and down magnetization. This increases the stability of the computational
states of the magnets but it is not the reason why the majority gate becomes more
reliable. The key parameter is the ratio of the height of the energy barrier
surrounding the local energy minimum around the magnetization in the $x$ direction, and
the strength of the signal produced by neighboring magnets. We ran our simulations 
for different sizes of the nanomagnets, but keeping a 2:1 aspect ratio and a thickness
of $6nm$. Figure \ref{Fig:MinErrorSize} shows these results.

We can see that the error probability decreases fast with size. The mechanism for this
behavior is the following. When we increase the size of the magnets following the 
prescription mentioned above, the depth of the local minimum increases, but this increase
is approximately a linear function of the length. On the other hand, the volume of the magnet
increase quadratically with the length (since we are keeping a fixed aspect ratio), and
hence the strength of the magnetic field generated by the magnets also increases quadratically.
In summary, the deeper local minimum does a better job stabilizing the magnet against 
thermal fluctuations, while the magnetic interaction grows faster, preventing the 
biaxial anisotropy from freezing the nanomagnets.
From these results we see that nanomagnets with size less that $200nm$ have too high 
gate error probability and thus cannot be used to build MQCA.

\section{Summary and discussion}
\label{conclusions}

The goal of this work was to estimate the characteristics
of an MQCA-based logic device, in particular the limits
that can be achieved in terms of minimum size, gate switching time, 
switching energy, and gate error probability. 
To this end we analyzed a simplified model 
in an effort to understand how these features are affected by
the basic parameters that characterize the MQCA. A reasonable requirement on 
the bit stability
of these devices naturally leads to a lower bound on the size
of the basic element of any MQCA. A nanomagnet must be at
least $20 nm$ long in one of its dimensions to prevent 
thermal fluctuations from inducing an error rate larger than
that of today's CMOS transistors. Furthermore, reducing this
size results in a rapidly degrading bit stability of the components,
making its applications in logic circuits less useful. 
Fault-tolerant design does not seem to help in this
situation, since any reduction in the size of the nanomagnets
will be offset by the increase in their number due to
the overhead usually accompanies fault-tolerant implementations.
Another way to push beyond this limit would be to work
at much lower temperatures, but that regime will not be practical
in the most common situations.

The lower bound on size also provides us with an estimate
 of switching for MQCA. After initialization of an MQCA, 
energy is dissipated when the magnetization of each nanomagnet
``rolls down'' the energy barrier until it reaches a minimum energy
configuration (like a ball rolling on curved surface in the presence
of friction.) Then, the energy dissipated by each nanomagnet 
is just the energy it had at the top of the barrier, and that is just
the height of the barrier. From the bit stability constraint we found that 
this height should be at least $1 eV$, and hence a MQCA could
in principle dissipate about $1 eV$ per nanomagnet. A logic gate such
as the majority gate requires only five nanomagnets, so we
could perform logic functions with a switching energy as low as
a few electron-volts. This is a big advantage of MQCA over CMOS
transistors, that requires several thousand electron-volt to operate
\cite{salahuddin2007}.
This is, however, only a theoretical limit, and it does not take into account
the practical difficulties of efficiently transferring such a small amount of energy
to each nanomagnet. 

To estimate the speed of MQCA logic gates we considered a very
simple model in which we approximated the nanomagnets by point
dipoles when computing their interaction, but included the
effects of geometrical and crystalline anisotropies through the computation of
the effective field. This approach is less sophisticated than the
micromagnetic simulations that have been used in the literature to study
similar systems, but our goal was not to obtain a very detailed picture
of the dynamics, but rather to have a good estimate of the fastest gate
time MQCA can achieve. Our model includes all the fundamental elements
of MQCA dynamics, and more refined simulations are likely to result 
in slower gate times. Using this simple model we found that 
the majority gate produces the required output in about $700 \, ps$,
which is slower than gate times expected from CMOS in the next few years.

Another obstacle for implementing MQCA-based logic has to do with
information transmission. In MQCA this is accomplished following
the same basic principles as logic. Chains of nanomagnets propagate
a signal through the dipole-dipole interaction. But the propagation
speed of this signal turns out to be around $100 \, m/sec$, which
is extremely slow when compared with the speed of electric signals
in a wire (typically around $10^7 \, m/sec$.) This is a huge disadvantage
for any MQCA scheme.

MQCA suffers from the problem that its nanomagnets are initialized in an unstable
state before the computation. Thermal fluctuation will push the nanomagnets
randomly into one of the stable states regardless of the value prescribed by the 
computation.
It has been proposed~\cite{carlton2008} that exploiting the
biaxial anisotropy of the material, can increase the robustness
of the MQCA initial state against thermal fluctuations, preventing
premature relaxation of the nanomagnets before the computation is 
complete. On the other hand, a strong biaxial anisotropy can 
completely freeze the dynamics, by trapping the magnetization
in the local energy minimum of the initial state. We simulated the 
behavior of the majority gate in the presence of thermal fluctuations
and analyzed the error rate of the majority gate for different
values of the biaxial anisotropy, in order to find what are the
optimal choices of the parameters. We found that for room temperature
operation ($T=300K$), the gate error rate has an impractically high value 
($ > 1\%$) for all sizes of nanomagnet smaller than 200nm. 
This seems to show that the biaxial
anisotropy approach may not be enough to solve the gate error rate problem 
and scale MQCA logic to smaller sizes at room temperature.  

\appendix

\section{Properties of the discretized stochastic LLG equations}
\label{num}

In this appendix we show some of the details of the numerical approach used to
solve the stochastic LLG equations in the presence of thermal fields. As mentioned
before, we follow essentially the approach presented in~\cite{d'aquino2006a}, 
that uses the midpoint rule to discretize the stochastic LLG equations. Here we
will show that introducing an extra term in the effective field that represents the
effects of the biaxial anisotropy does not change the two main properties of this
technique, namely the unconditional preservation of the magnetization magnitude
and the consistency of the evolution of the free energy. 

The stochastic LLG equations take the form
\be
\label{stochasticLLG}
d \mathbf{m}^{(i)} = -\mathbf{m}^{(i)} \times 
\mathbf{h}^{(i)}_{eff} \,dt -\mathbf{m}^{(i)} \times \nu \, 
d\mathbf{W}^{(i)} + \alpha\, 
 \mathbf{m}^{(i)} \times d \mathbf{m}^{(i)},
\ee
where $\mathbf{h}^{(i)}_{eff}$ includes the biaxial term.
Applying the midpoint method corresponds to the following replacements:
\begin{eqnarray}
d \mathbf{m}^{(i)} & \longrightarrow &  \left( \mathbf{m}^{(i)}_{n+1} - \mathbf{m}^{(i)}_n \right)\\
\mathbf{m}^{(i)} & \longrightarrow & \left(\frac{\mathbf{m}^{(i)}_{n+1} + \mathbf{m}^{(i)}_n}{2}\right) \\
\mathbf{h}^{(i)}_{eff} (\mathbf{m}^{(i)},t_n) & \longrightarrow & \mathbf{h}^{(i)}_{eff} 
\left(\frac{\mathbf{m}^{(i)}_{n+1} + \mathbf{m}^{(i)}_n}{2},t_n + \frac{\Delta t}{2}\right) \\
d\mathbf{W}^{(i)} & \longrightarrow & \left(\mathbf{W}^{(i)}_{n+1} - \mathbf{W}^{(i)}_n\right)
\end{eqnarray}
that result in the discretized stochastic LLG equations
\begin{eqnarray}
\label{discSLLG}
\left( \mathbf{m}^{(i)}_{n+1} - \mathbf{m}^{(i)}_n \right) & = & -\left(\frac{\mathbf{m}^{(i)}_{n+1} 
+ \mathbf{m}^{(i)}_n}{2}\right) \times 
\mathbf{h}^{(i)}_{eff}\left(\frac{\mathbf{m}^{(i)}_{n+1} + \mathbf{m}^{(i)}_n}{2},t_n 
+ \frac{\Delta t}{2}\right)  \Delta t  - \\ \nonumber
& & \ \ - \left(\frac{\mathbf{m}^{(i)}_{n+1} + \mathbf{m}^{(i)}_n}{2}\right) \times \nu \, 
\left(\mathbf{W}^{(i)}_{n+1} - \mathbf{W}^{(i)}_n\right)  + \\ \nonumber
& & \ \ + \alpha\, \left(\frac{\mathbf{m}^{(i)}_{n+1} 
+ \mathbf{m}^{(i)}_n}{2}\right) \times \left( \mathbf{m}^{(i)}_{n+1} - \mathbf{m}^{(i)}_n \right).
\end{eqnarray}
Since every term on the RHS is of the form $\left(\mathbf{m}^{(i)}_{n+1} + \mathbf{m}^{(i)}_n\right) 
\times \mathbf{V}$, it is clear that the RHS vanishes when scalar multiplied by 
$\left(\mathbf{m}^{(i)}_{n+1} + \mathbf{m}^{(i)}_n\right)$, while the LHS becomes 
$(|\mathbf{m}^{(i)}_{n+1}|^2 - |\mathbf{m}^{(i)}_n|^2)$, and so we have that
\be
\label{normpreserve}
|\mathbf{m}^{(i)}_{n+1}|^2 = |\mathbf{m}^{(i)}_n|^2,
\ee
which means the midpoint method unconditionally preserves the magnitude of the magnetization. 
Note that the form of the term added to the effective field does not affect this property, 
since the corresponding term on the RHS is still of the form 
$\left(\mathbf{m}^{(i)}_{n+1} + \mathbf{m}^{(i)}_n\right) \times \mathbf{V}$.

Another property that the discretized stochastic LLG equations presented in~\cite{d'aquino2006a} 
have is that the change in the discretized free energy is bounded by the work performed by the thermal
fields on the magnetization for any finite value of the increment $\Delta t$. Their proof of this
fact relies on the particular form of the effective field, namely that the free energy is an 
at most quadratic polynomial function of the magnetization. Even though when we add the 
biaxial anisotropy term the free energy has a term of degree 4, the result still holds as we show below.
First we write the free energy $g(\mathbf{m})$
\begin{eqnarray}
\label{freeenergy}
g(\mathbf{m}) &=& \frac{1}{2} \sum_i \mathbf{m}^{(i)} \cdot {\cal N} \cdot \mathbf{m}^{(i)} 
- \frac{1}{2} \sum_i \sum_{j \neq i} \mathbf{m}^{(i)} \cdot  C^{(ij)} \cdot \mathbf{m}^{(j)} - \\ \nonumber  
& & \ \ - \sum_i \mathbf{h}_{ext}^{(i)} \cdot \mathbf{m}^{(i)} - \frac{2 K_2}{\mu_0 M_s^2} (m_x (1-m_x^2) 
\hat{\mathbf{x}} + m_y (1-m_y^2) \hat{\mathbf{y}} +
m_z (1-m_z^2) \hat{\mathbf{z}}),
\end{eqnarray}
where $\cal{N}$ is the demagnetization tensor, which includes a term corresponding to the uniaxial 
anisotropy, $C^{(ij)}$ is the matrix that encodes the dipole-dipole interaction between  nanomagnets 
(and it is symmetric with respect to $i$ and $j$.) We want to compute $g_{n+1} - g_n$, 
where $g_n = g(\mathbf{m}_n)$. Clearly, this will give us an expression in powers of 
$\delta \mathbf{m}^{(i)} = (\mathbf{m}^{(i)}_{n+1} - \mathbf{m}^{(i)}_{n})$. We will keep terms up to order 
$(\delta \mathbf{m}^{(i)})^2$, since from the LLG equations we can see that 
$\delta \mathbf{m}^{(i)}$ is proportional to $ \left(\mathbf{W}^{(i)}_{n+1} - \mathbf{W}^{(i)}_n\right)$,
and $\left(\mathbf{W}^{(i)}_{n+1} - \mathbf{W}^{(i)}_n\right)^2$ is of order $\Delta t$ . 
With that in mind, after some algebra we get
\begin{eqnarray}
g_{n+1} - g_n & \simeq & \sum_i \left(\mathbf{m}^{(i)} \cdot {\cal N}  
-  \sum_{j \neq i} \mathbf{m}^{(j)} \cdot C^{(ij)} \cdot \mathbf{m}^{(j)}  
- \mathbf{h}_{ext}^{(i)}  - \mathbf{h}_{biaxial}^{(i)}\right) \cdot \delta \mathbf{m}^{(i)} + \\ \nonumber
& & \ \ \  + \frac{1}{2} \sum_i \delta \mathbf{m}^{(i)} \cdot {\cal N} \cdot  \delta 
\mathbf{m}^{(i)} - \frac{1}{2} \sum_i \sum_{j \neq i} \delta \mathbf{m}^{(i)} \cdot C^{(ij)} 
\cdot \mathbf{m}^{(j)} - \\ \nonumber
& & \ \ \ - \frac{K_2}{\mu_0 M_s^2} \sum_i \delta \mathbf{m}^{(i)} \cdot {\cal D} 
\cdot \delta \mathbf{m}^{(i)},
\end{eqnarray}
where ${\cal D}= \mathbf{1} + 2 \mathbf{m}_n^{(i)} \mathbf{m}_n^{(i)\,\mathbf{T}} - 3\, 
diag \left((m_x^{(i)})_n^2,(m_y^{(i)})_n^2,(m_z^{(i)})_n^2\right)$. Note that the term multiplying 
$\delta \mathbf{m}^{(i)}$ in the first sum is exactly $-\mathbf{h}_{eff}^{(i)}(\mathbf{m}_n)$ 
(as it should be). Now we go back to the discretized LLG equations that have the form
\be
\mathbf{m}^{(i)}_{n+1} - \mathbf{m}^{(i)}_{n} = - \mathbf{m}^{(i)}_{n+\frac{1}{2}} 
\times \left(\mathbf{h}^{(i)}_{eff} (\mathbf{m}^{(i)}_{n+\frac{1}{2}},t_n + \frac{\Delta t}{2}) 
\Delta t + \nu \, \left(\mathbf{W}^{(i)}_{n+1} - \mathbf{W}^{(i)}_n\right)  + \alpha \, 
\left(\mathbf{m}^{(i)}_{n+1} - \mathbf{m}^{(i)}_n\right) \right)
\ee
where now $\mathbf{h}^{(i)}_{eff}$ also includes the biaxial term. This equation is of 
the form $\mathbf{m}^{(i)}_{n+1} - \mathbf{m}^{(i)}_{n} = - \mathbf{m}^{(i)}_{n+\frac{1}{2}} \times \mathbf{A}$, 
and so if we scalar multiply both sides by $\mathbf{A}$, the RHS vanishes and we get
\be
\mathbf{h}^{(i)}_{eff}\left(\mathbf{m}^{(i)}_{n+\frac{1}{2}},t_n + \frac{\Delta t}{2}\right)   \cdot 
\delta \mathbf{m}^{(i)}\, \Delta t +\nu \, \left(\mathbf{W}^{(i)}_{n+1} - \mathbf{W}^{(i)}_n\right) 
\cdot \delta \mathbf{m}^{(i)}  = \alpha |\delta \mathbf{m}^{(i)}|^2.
\ee
Now we write $\mathbf{h}^{(i)}_{eff}\left(\mathbf{m}^{(i)}_{n+\frac{1}{2}},t_n + \frac{\Delta t}{2}\right)$ 
in terms of $\mathbf{h}^{(i)}_{eff}\left(\mathbf{m}^{(i)}_{n}\right)$ 
(we drop the time, since the field does not have an explicit time dependence). 
After some more algebra, we get
\begin{eqnarray}
\mathbf{h}^{(i)}_{eff}\left(\mathbf{m}^{(i)}_{n+\frac{1}{2}}\right)  \cdot \delta \mathbf{m}^{(i)}\Delta t  
& = & \mathbf{h}^{(i)}_{eff}\left(\mathbf{m}^{(i)}_{n}\right) \Delta t  + \Delta t  \sum_{j \neq i} 
\delta \mathbf{m}^{(j)} \cdot C^{(ij)} \cdot \delta \mathbf{m}^{(i)} -  \\ \nonumber
& & \ \ - \Delta t \,\delta \mathbf{m}^{(i)} 
\cdot \left( \frac{K_2}{\mu_0 M_s^2} (3 M^{(i)} - \mathbf{1}) -\frac{1}{2} {\cal N} \right) \cdot \delta 
\mathbf{m}^{(i)}
\end{eqnarray}
with $M^{(i)} = diag\left((m_x^{(i)})_n^2,(m_x^{(i)})_n^2,(m_x^{(i)})_n^2\right)$. Now, using this in the 
expression we computed for $g_{n+1} - g_n$, and doing even more algebra, we arrive to
\be
g_{n+1} - g_n = \frac{\nu}{\Delta t} \sum_i \left( \mathbf{W}^{(i)}_{n+1} - \mathbf{W}^{(i)}_n\right) 
\cdot \delta \mathbf{m}^{(i)} - \sum_i \delta \mathbf{m}^{(j)} \cdot {\cal M}^{(i)}(\mathbf{m}_n) 
\cdot \delta \mathbf{m}^{(i)}, 
\ee
with $ {\cal M}^{(i)}(\mathbf{m}_n) = \frac{\alpha}{\Delta t} \mathbf{1} + \frac{2 K_2}{\mu_0 M_s^2} \, 
\mathbf{m}^{(i)}_n \mathbf{m}^{(i)\, \mathbf{T}}_n$. Hence, $ {\cal M}^{(i)}(\mathbf{m}_n)$ is positive semidefinite, 
and so we have finally
\be
g_{n+1} - g_n \leq \nu (\mathbf{m}^{(i)}_{n+1} - \mathbf{m}^{(i)}_{n})\cdot \frac{\left(\mathbf{W}^{(i)}_{n+1} 
- \mathbf{W}^{(i)}_n\right)}{\Delta t},
\ee
which shows that the change in the discretized free energy is always less than the work done by the 
stochastic field during the time interval $\Delta t$.

\bibliography{mqca5}

\end{document}